\documentclass[
reprint,
superscriptaddress,
 amsmath,amssymb,
 aps,
 prl,
floatfix,
]{revtex4-1}

\usepackage{graphicx}
\usepackage{dcolumn}
\usepackage{bm}
\usepackage{ifpdf}
\usepackage[usenames,dvipsnames]{color}
\usepackage{hyperref}
\usepackage[all]{hypcap}
\hypersetup{urlcolor=Blue, citecolor=Blue,linkcolor=Blue,colorlinks=true}

\begin{document}

\title{Anomalous Heating and Plasmoid Formation in a\protect\\Driven Magnetic Reconnection Experiment}

\author{J. D. Hare}
\email{jdhare@imperial.ac.uk}
\author{L. Suttle}
\author{S. V. Lebedev}
\email{s.lebedev@imperial.ac.uk}
\affiliation{Blackett Laboratory, Imperial College, London, SW7 2AZ, United Kingdom}
\author{N. F. Loureiro}
\affiliation{Plasma Science and Fusion Center, Massachusetts Institute of Technology, Cambridge MA 02139, USA}
\author{A. Ciardi}
\affiliation{Sorbonne Universit\'{e}s, UPMC Univ Paris 06, Observatoire de Paris, PSL Research University, CNRS, UMR 8112, LERMA, F-75005, Paris, France}
\author{G. C. Burdiak}
\author{J. P. Chittenden}
\author{T. Clayson}
\author{C. Garcia}
\author{N. Niasse}
\author{T. Robinson}
\author{R. A. Smith}
\author{N. Stuart}
\author{F. Suzuki-Vidal}
\author{G. F. Swadling}
\altaffiliation[Current address: ]{Lawrence Livermore National Laboratory, California 94550, USA}
\affiliation{Blackett Laboratory, Imperial College, London, SW7 2AZ, United Kingdom}
\author{J. Ma}
\affiliation{Northwest Institute of Nuclear Technology, Xi'an 710024, China}
\author{J. Wu}
\affiliation{Xi'an Jiaotong University, Shaanxi 710049, China}
\author{Q. Yang}
\affiliation{Institute of Fluid Physics, China Academy of Engineering Physics, Mianyang 621900, China}


\begin{abstract}
We present a detailed study of magnetic reconnection in a quasi-two-dimensional pulsed-power driven laboratory experiment.
Oppositely directed magnetic fields (\(B=3\) T), advected by supersonic, sub-Alfv\'enic carbon plasma flows (\(V_{in}=50\) km/s), are brought together and mutually annihilate inside a thin current layer (\(\delta=0.6\) mm).  
Temporally and spatially resolved optical diagnostics, including interferometry, Faraday rotation imaging and Thomson scattering, allow us to determine the structure and dynamics of this layer, the nature of the inflows and outflows and the detailed energy partition during the reconnection process.
We measure high electron and ion temperatures (\(T_e=100\) eV, \(T_i=600\) eV), far in excess of what can be attributed to classical (Spitzer) resistive and viscous dissipation.
We observe the repeated formation and ejection of plasmoids, consistent with the predictions from semi-collisional plasmoid theory.
\end{abstract}

\maketitle

Magnetic reconnection is the rapid change of magnetic field topology in a plasma, accompanied by bulk heating and particle acceleration \cite{Zweibel2009, Yamada2010}.
Reconnection is a ubiquitous process which occurs across a vast region of parameter space, including the collisionless plasmas at the heliopause \cite{Opher2011} and the dense, hot plasmas deep in the solar convection zone \cite{Fan2004, Ryutov2015a}. 
Understanding of magnetic reconnection has improved over the years thanks to dedicated laboratory experiments. 
In facilities like MRX \cite{Ji2004, Yamada2015a, Yamada2016} and TREX \cite{Olson2016} the magnetic energy is much larger than the other plasma energy components.
In contrast, laser-driven high energy density plasma (HEDP) experiments are strongly driven --- the kinetic and thermal energies are much larger than the magnetic energy \cite{Nilson2006, Fiksel2014b}, and reconnection heating is small \cite{Rosenberg2012}.
\begin{figure}[t]
\label{fig:setup}
\ifpdf
	\includegraphics[scale=1]{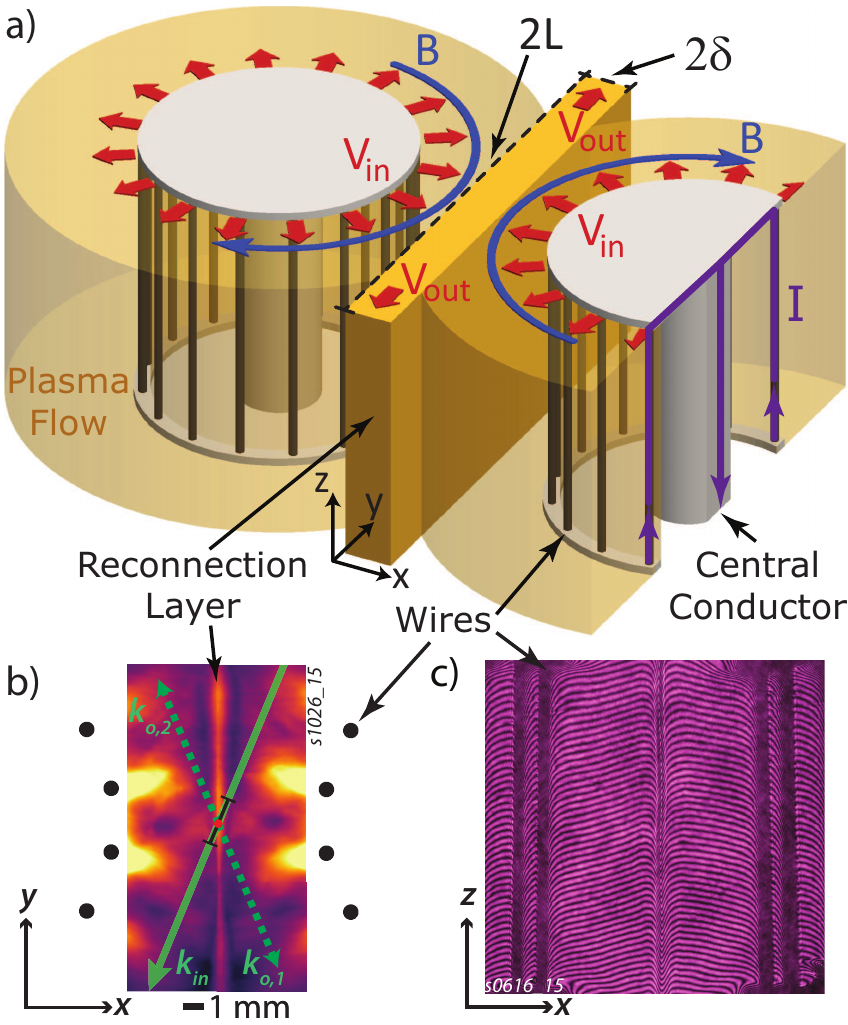} 
\else
	\includegraphics[scale=1]{reconnection_paper-01.eps} 
\fi
\centering
\caption{a) Experimental setup with geometry of reconnection layer. The cutaway on the right array shows the current path. b) Top view with density map (taken at \(t=272\) ns after current start) and Thomson scattering vectors. c) Side view interferogram.}
\end{figure}

In this letter we present experimental studies of HEDP magnetic reconnection driven by a new pulsed-power platform.
The reconnection layer was created by the interaction of magnetised plasma flows in a quasi-2D geometry, which we studied using high resolution, non-perturbative measurements of the temperature, flow velocity, electron density and magnetic field in the reconnection layer. 
The colliding plasma flows were supersonic (\(M_s\sim1.6\)) but sub-Alfv\'enic (\(M_A\sim 0.7\)), and therefore the thermal and dynamic plasma betas (ratio of the thermal or ram pressure to the magnetic pressure) are close to unity (\(\beta_{th}\sim0.7\), \(\beta_{dyn}\sim0.9\)).
These parameters are significantly different to those found both in magnetically driven experiments, such as MRX, and in laser driven experiments, and we believe our experiments are the first to make a detailed study of this regime.
We observed the formation of a reconnection layer with an aspect ratio of \(L/\delta>10\), which existed for at least ten hydrodynamic flow times \(\delta/V_{in}\), where \(L\) is the layer half-length and \(\delta\) is the layer half-width (\hyperref[fig:setup]{Fig. 1a}). 
The annihilation of the magnetic flux caused strong plasma heating in the reconnection layer (\(T_i\approx600\) eV, \(\bar{Z}T_e\approx600\) with \(T_e\approx100\) eV in a carbon plasma with average ionisation \(\bar{Z}\approx6\)). 
The ion temperature in the layer was more than five times greater than the kinetic energy of the incoming ions, consistent with strong reconnection heating.
Although we show that there is a balance between the measured power flow into and out of the reconnection layer, the mechanism which converts magnetic energy to thermal energy is currently unclear, as the time-scales for viscous or resistive heating are too long to heat the ions or the electrons.

The experimental setup is illustrated in \hyperref[fig:setup]{Fig. 1a}, which is similar to the setup in \cite{Suttle2016}, except that the plasma was made from carbon rather than aluminium.
This carbon plasma was in a different region of parameter space to \cite{Suttle2016}, with sub-Alfv\'enic flows and a significantly reduced rate of radiative cooling. 
Reduced cooling allowed this plasma to attain higher electron temperatures than in \cite{Suttle2016}, and hence a higher Lundquist number of around 120 (\(S=L V_A/\eta\propto{}T_e^{3/2}\), where \(V_A\) is the Alfv\'{e}n velocity and \(\eta\) is the magnetic diffusivity).
The interacting plasma flows were produced by the ablation of material \cite{Lebedev2001a} from two `inverse' cylindrical carbon arrays \cite{Harvey-Thompson2009} placed side by side and driven in parallel by a 1.4 MA, 500 ns current pulse from the MAGPIE generator \cite{Mitchell1996}. 
The current was divided equally between the two arrays --- each array consisted of 16 parallel carbon wires (400 \(\mu\)m diameter, 16 mm tall) equally spaced around a circle (16 mm diameter), concentric to a central conductor, and 27 mm apart from the other array (field line curvature at mid-plane \(R_c=13.5\) mm).
The global azimuthal magnetic field accelerates the ablated plasma outwards. 
Some of the drive current switches into the plasma surrounding the wires, which means that a fraction of the global magnetic field is advected by the plasma flows into the initially field-free region surrounding the arrays \cite{Chittenden2004, Greenly2009}.
A continuous flow of magnetised plasma \cite{Lebedev2014} was delivered for the duration of the drive current \cite{Swadling2016}, and reconnection occurred when the embedded anti-parallel magnetic fields met at the mid-plane.

The reconnection layer was highly uniform, as can be seen from laser probing images in \hyperref[fig:setup]{Fig. 1b, c}.  
An electron density map in the reconnection (\(x,y\)) plane is shown in \hyperref[fig:setup]{Fig. 1b}, demonstrating the formation of the elongated layer. 
The reconnection layer was also uniform in the out of plane (\(z\)) direction, as seen in the side-on ((\(x,z\)) plane) laser interferogram (\hyperref[fig:setup]{Fig. 1c}), which justifies treating the system as quasi-2D for our analysis.

\begin{figure}[t]
\label{fig:density}
\ifpdf
\includegraphics[scale=1]{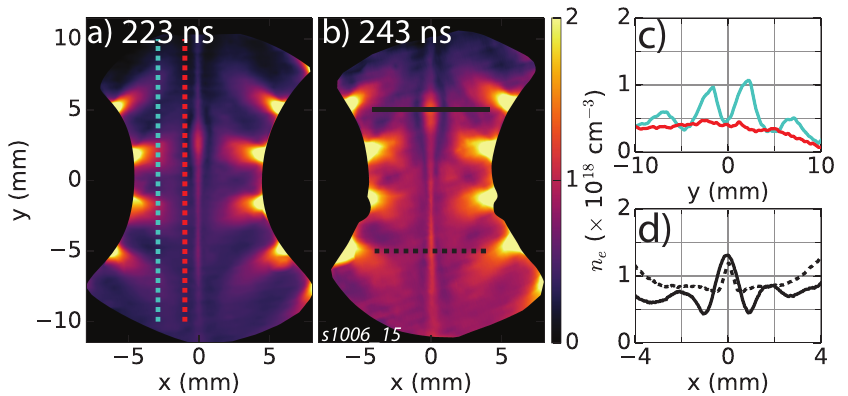} 
\else
\includegraphics[scale=1]{reconnection_paper-02.eps} 
\fi
\centering
\caption{Electron density maps from laser interferometry, both from the same shot. a) At 223 ns after current start. b) At 243 ns after current start. In both a) and b) there is an obvious region of enhanced density (a `plasmoid') inside the reconnection layer. c) Lineouts of electron density, \(x\) positions shown in a). d) Lineouts of electron density across the reconnection layer, \(y\) positions shown in b).}
\end{figure}

Quantitative measurements of the plasma parameters in the reconnection layer were made using interferometry, Faraday rotation polarimetry and Thomson scattering diagnostics \cite{Swadling2014a}.
The experimental results discussed in this paper are highly reproducible. 
The electron density distribution in the reconnection (\(x,y\)) plane was measured using a two-frame Mach-Zehnder laser interferometry system (532 nm and 355 nm, same optical path, 0.4 ns pulse length). 
The interferometry analysis was performed as in \cite{Swadling2014a}, and \hyperref[fig:density]{Fig. 2a} and \hyperref[fig:density]{Fig. 2b} show typical electron density maps obtained in the same experiment 20 ns apart.
An elongated reconnection layer had formed by \(t\lesssim180\) ns, and at \(t=223\) ns (\hyperref[fig:density]{Fig. 2a}) the outflows extended for the entire field of view of the diagnostic (22 mm), with a layer half-width of \(\delta\approx0.6\) mm.
The layer was formed by the interaction of radially diverging flows produced by the two arrays of discrete wires.
Close to the arrays, the density was modulated by the discrete number of wires, but this modulation was significantly reduced as the flows approached the mid-plane. \hyperref[fig:density]{Fig. 2c} shows electron density profiles \(n_e(y)\) measured along two lines indicated in \hyperref[fig:density]{Fig. 2a}.
At \(x=-3\) mm from the mid-plane \(n_{e,max}/n_{e,min}\sim3\), while at \(x=-1\) mm the density modulations were negligible. 
Typical electron densities in the flow just outside of the layer were \(n_e=0.3-0.8\times10^{18}\) cm\(^{-3}\).

\hyperref[fig:density]{Fig. 2a} and \hyperref[fig:density]{2b} show the presence of a localised elliptical region of enhanced electron density, which we will call a plasmoid. 
The plasmoid was seen at \(y=2.5\) mm at \(t=223\) ns, and at \(y=5.0\) mm at \(t=243\) ns, which corresponds to a propagation speed of \(V_y\approx130\) km/s. 
The presence of plasmoids was reproducible between experiments, but the time and location at which the plasmoids appeared was stochastic.
There was a marked depletion of electron density just outside of the layer at \(x\approx0.7\) mm, visible in \hyperref[fig:density]{Fig. 2a} and \hyperref[fig:density]{2b}, in the lineouts in \hyperref[fig:density]{Fig. 2d}, and especially evident around plasmoids.

\begin{figure}[t]
\label{fig:faraday}
\ifpdf
\includegraphics[scale=1]{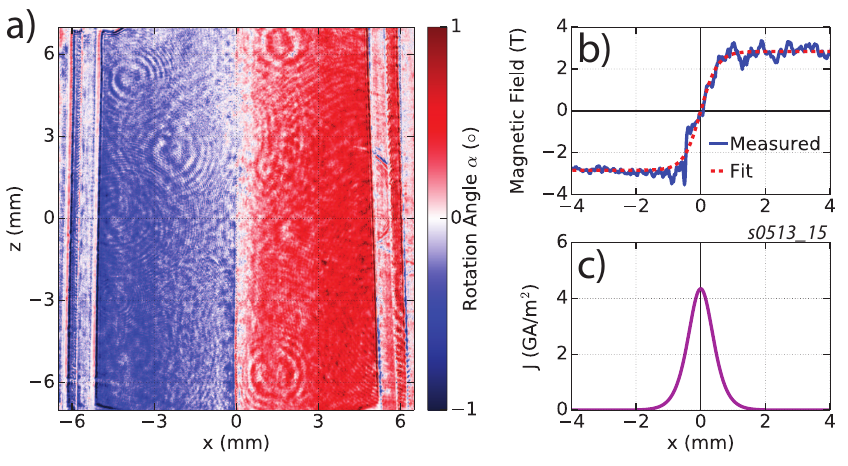} 
\else
\includegraphics[scale=1]{reconnection_paper-03.eps} 
\fi
\centering
\caption{Data from Faraday effect polarimetry, taken at \(t=251\) ns after current start. a) Rotation angle of the linear polarisation of the laser beam passing in the \(y\) direction. b) Measured magnetic field profile (blue) and Harris sheet fit (red). c) Electric current density calculated from the Harris sheet fit.}
\end{figure}
The spatial distribution of the reconnecting magnetic field was measured using a Faraday-effect laser-polarimetry diagnostic \cite{Swadling2014a}. 
The probing was in the \(y\)-direction, producing  images in the (\(x,z\)) plane, as in \hyperref[fig:setup]{Fig. 1c}.
\hyperref[fig:faraday]{Fig. 3a} is a polarogram, which shows the angle of rotation of the linear polarisation of the probing laser beam (1053 nm, 5 J, 1 ns), obtained at \(t=251\) ns, 8 ns after the electron density map in \hyperref[fig:density]{Fig. 2b}. 
The rotation angle was fairly uniform in the \(z\) direction, and had opposite signs on opposite sides of the mid-plane, with a maximum absolute value of \(\sim1^{\circ}\).
To determine the line averaged magnetic field \(B_y(x)\), we used this polarogram and a line integrated electron density map, which was obtained by interferometry (\hyperref[fig:setup]{Fig. 1c}) using the same probing laser beam as the polarimetry \cite{Swadling2014a}. 

\hyperref[fig:faraday]{Fig. 3b} shows the profile of \(B_y(x)\) (blue line) averaged in the \(z\)-direction over 1.5 mm around \(z=0\) mm. 
The measured magnetic field is well approximated by the Harris profile \(B_y(x)=B_0\tanh(x/\delta)\) (red dashed line, \cite{Harris1962}) with \(B_0=3\) T, and we find the layer half width is \(\delta=0.6\) mm, consistent with the electron density measurements. 
Overall, the measured structure of the magnetic field is consistent with annihilation of the magnetic flux in the reconnection layer, and there is no evidence of flux pileup outside the reconnection layer.
We observed two additional signatures of magnetic reconnection: strong heating of the plasma and fast outflows along the reconnection layer with velocities exceeding \(V_A\).
The plasma temperature and flow velocities were measured using a Thomson scattering (TS) diagnostic, which recorded the ion feature of the scattering spectra simultaneously from fourteen spatial locations along the probing laser beam (\hyperref[fig:setup]{Fig. 1b}). 
The focused laser beam (532 nm, 3 J, 8 ns pulse length, beam width \(\sim 100\) \(\mu\)m) propagated in the (\(x,y\)) plane through the centre of the reconnection layer, and the scattered light was collected in the same plane, at angles of \(45^{\circ}\) and \(135^{\circ}\) to the laser beam (\(\textbf{k}_{o,1}\) \& \(\textbf{k}_{o,2}\) respectively), as shown in \hyperref[fig:setup]{Fig. 1b} (see \cite{Swadling2014a} for more details). 
The two resultant scattering vectors (\(\textbf{k}_{S,j}=\textbf{k}_{o,j}-\textbf{k}_{in}\)) give Doppler shifted spectra sensitive to velocity components (\(\delta \omega_j=\textbf{V}\cdot\textbf{k}_{S,j}\)) in the \(x\) or \(y\) directions only (\(\textbf{k}_{o,1}\) \& \(\textbf{k}_{o,2}\) respectively), and the spectra were fit using theoretical form factors to infer velocity and temperature \cite{Swadling2014, Froula2011}.
\begin{figure}[t]
	\label{fig:thomson}
	\ifpdf
	\includegraphics[scale=1]{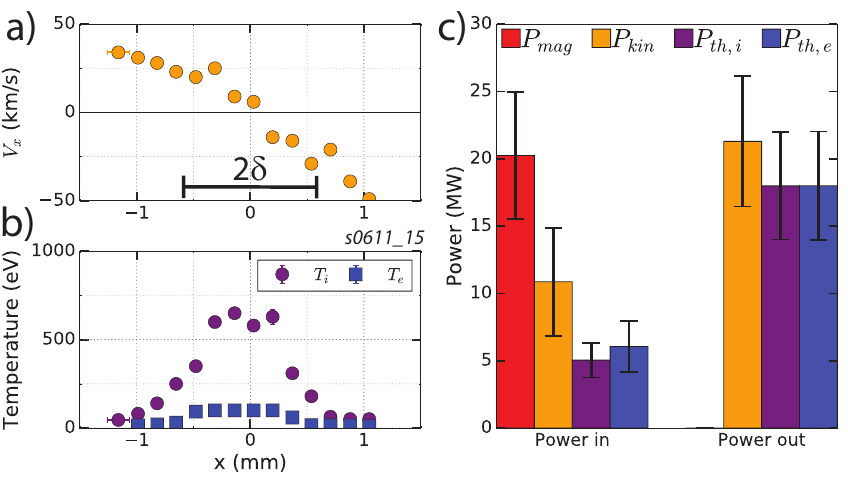} 
	\else
	\includegraphics[scale=1]{reconnection_paper-04.eps} 
	\fi
	\centering
	\caption{Thomson Scattering measurements taken at \(t=232\) ns after current start. a) Inflow velocity. b) Electron and ion temperatures. Spatial error-bar shown for first data point. c) Calculated power flow into and out of the reconnection layer.}
\end{figure}

Typical results of the TS measurements are shown in \hyperref[fig:thomson]{Fig. 4}, where we present spatial profiles of the inflow velocity (\(V_x\), \hyperref[fig:thomson]{Fig. 4a}) and of the electron and ion temperatures (\hyperref[fig:thomson]{Fig. 4b}). 
The scattering volumes (200 \(\mu\)m spot size) were separated by 420 \(\mu\)m along a chord which passed through the origin at an angle of \(22.5^{\circ}\) to the \(y\) axis, giving a 5.9 mm field of view (\(\Delta{}x=2.3\) mm, \(\Delta{}y=5.5\) mm). 
Outside of the reconnection layer (\(x\approx 1\) mm) the flow was predominantly perpendicular to the layer (\(V_x=50\) km/s), and the same \(V_x\) was also measured further upstream, at \(x\approx\) 3 mm.
Inside the reconnection layer, \hyperref[fig:thomson]{Fig. 4a} shows that the inflow velocity gradually decreased from \(V_x\approx\pm50\) km/s at \(\lvert x\rvert=1\) mm to zero in the centre of the layer. 
Over the same spatial scale there was a significant increase in the electron and ion temperatures (\hyperref[fig:thomson]{Fig. 4b}).
In the upstream flow \(T_i\leq{}50\) eV and \(\bar{Z}T_e\approx 60\) eV (corresponding to \(T_e=15\) eV for \(\bar{Z}= 4\), determined by a non-local thermodynamic equilibrium (nLTE) ionisation model \cite{Chittenden2016}).
In the reconnection layer the temperatures were significantly higher, reaching \(T_i\approx 600\) eV and \(\bar{Z}T_e\approx 600\) eV (\(T_e=100\) eV, \(\bar{Z}=6\) in nLTE).
The ion temperature measured in the reconnection layer was much larger than the kinetic energy of the ions (\(E_i=m_{i}V^2/2=150\) eV) entering the layer, and so clearly the measured ion temperature cannot be explained by the thermalisation of the inflow kinetic energy alone.

The outflow velocity \(V_y\) along the layer was measured in a different TS scattering geometry.
The laser passed along  the reconnection layer, and the scattered light was collected in the \(\boldsymbol{\hat{\textbf{z}}}\) direction, such that the velocity measured was \(V=(V_y+V_z)/\sqrt{2}\).
In 2D geometry \(V_z\) is zero, and so we infer \(V_y=130\) km/s at \(y=5\) mm, consistent with the plasmoid propagation velocity inferred from \hyperref[fig:density]{Fig. 2}.

\begin{table}[h]
	\label{tab:params}
	\centering
	\caption{Plasma parameters in the inflowing plasma and reconnection layer.}
	\begin{tabular}{|c|c|r|r|r|r|r|r|r|}
		\hline
		\rule{0pt}{2.5ex}
		Parameter & \(n_e\)            & \(\bar{Z}\)& \(V_x \left(V_y\right)\) & \(B_y\) & \(T_i\) & \(T_e\)     & \(c/\omega_{pi}\) & \(\lambda_{ii}\)\\ 
		\textit{Units}     & (cm\(^{-3}\))       & & (km/s)                  & (T) & (eV)  & (eV)      & (\(\mu\)m)          & (\(\mu\)m) \\ 
		\hline
		Inflow    & \(3\times10^{17}\) &4& 50\hphantom{)}                      & 3     & 50 & 15 & 700               & 3                 \rule{0pt}{3ex}  \\ \hline
		Layer     & \(6\times10^{17}\) &6& (130)                &    -   & 600     &100         & 400               & 30 \rule{0pt}{3ex}  \\ 
		\hline
	\end{tabular}
\end{table}

The measured and derived plasma parameters relevant to reconnection are summarised in Table \ref{tab:params}. 
We observed the formation of a reconnection layer which existed for much longer (\(>\)200 ns) than the characteristic hydrodynamic time (\(\delta/V_{in}\approx12\) ns).
Using the measured plas ma parameters, we find that the thermal pressure in the layer was balanced by  equal contributions from the magnetic and ram pressures in the flow.
The magnetic field profile is well approximated by the Harris model, consistent with the annihilation of magnetic flux inside the reconnection layer, but two surprising results warrant further discussion:

1. The inflow velocity (\(V_x=50\) km/s), imposed by the large dynamic beta of the reconnecting flows, is much faster than the standard Sweet-Parker \cite{Parker1957, Sweet1958} model predicts (\(V_A\approx70\) km/s, \(S\)=120 \footnote{We use (\(L=R_c/2\approx7\) mm), half of the radius of magnetic field line curvature at the mid-plane.},  \(V_A/S^{1/2}\approx7\) km/s) and the outflows are significantly super-Alfv\'enic. 
These velocities are however consistent with the generalised Sweet-Parker model of Ji et al. \cite{Ji1999}, which includes compressibility effects and the difference in pressure between the upstream and downstream regions.
In our experiments the outflows expand into the vacuum and the predicted outflow speed (\cite{Ji1999}, eqn. 6) is
\begin{equation}
V_y=\sqrt{V_A^2+2C_{i, A}^2}= 140 \pm 4\,\textrm{km/s},
\end{equation}
where \(C_{i,A}=\sqrt{(\bar{Z}T_e+T_i)/m_i}\). This velocity closely agrees with TS measurements of 130 km/s.
The inflow speed is predicted to be (\cite{Ji1999}, eqn. 5, modified to account for ionisation inside the layer):
\begin{equation}
\label{eqn:compressibility}
V_x=\frac{\delta}{L}\left( V_y\frac{n_2}{n_1}+\frac{L}{n_1}\frac{\partial n_2}{\partial t}\right)=31\pm 4\, \textrm{km/s},
\end{equation}
where \(n_1\) is the ion density at the edge of the layer (\(x=\pm0.6\) mm) and \(n_2\) is the ion density at the centre of the layer (\(x=0\) mm). We calculate the ion densities using \(n_e\) from \hyperref[fig:density]{Fig. 2a} and \(\bar{Z}\) from TS, and we estimate \(\partial{}n_2/\partial{}t\) using electron densities measured in the same experiment with \(\Delta{}t=20\) ns, significantly less than the outflow transit time (\hyperref[fig:density]{Fig. 2a \& b}).
The velocity predicted by eqn. \ref{eqn:compressibility} is close to the measured velocity.

2. Both the electrons and ions were heated significantly during the reconnection process.
The overall power balance is shown in \hyperref[fig:thomson]{Fig. 4c}, which shows agreement within experimental error between the power into and out of the reconnection layer.
The powers are calculated by multiplying each energy density in the inflow or outflow regions (\(E_{mag}=B^2/2\mu_0\), \(E_{kin}=n_{i}m_i V^2/2\), \(E_{th,\alpha}=3k_{B}n_{\alpha}T_{\alpha}/2\)) by \(LV_{x}h\) (inflow) or \(\delta{}V_{y}h\) (outflow), where \(h=16\) mm is the height of the reconnection layer.
The overall power balance in \hyperref[fig:thomson]{Fig. 4c} suggests that in the outflow \(P_{mag}\) is negligible within the experimental uncertainty of the other energy components, which is consistent with the (collisional) Sweet-Parker model, \(P_{mag,out}=P_{mag,in}/S\approx0.01P_{mag,in}\) 
--- this is unlike collisionless reconnection, where detailed studies have shown the outflow magnetic energy to be significant \cite{Yamada2015a, Yamada2016}.

From \hyperref[fig:thomson]{Fig. 4c} it is clear that the annihilation of the magnetic field is the primary source of heating and acceleration for the electrons and ions, but the mechanism for this energy transfer is unclear.
 We can calculate the time scale for viscous heating of the ions as \(\tau_{visc}=800\) ns \cite{Hsu2001a}, using \(V_y=0\) km/s outside the layer (\(x=\pm0.6\) mm), and assuming that the ions are heated from 50 eV to 600 eV by viscous heating alone.
The time scale for significant viscous ion heating to occur is too long for our experiment, and so the ion heating is anomalous. 
The electrons are also anomalously heated: the expected electron temperature can be estimated from the Ohmic heating alone, because the other terms in the energy equation, such as radiative cooling, ion-electron energy exchange and parallel heat conduction \footnote{\(\tau_{rad}\approx600\) ns, calculated using an nLTE model \cite{Chittenden2016}, \(\tau_{E, ei}\approx250\) ns \cite{Ryutov2015a}, \(\tau_{\parallel}\approx1000\) ns}, are not significant on the experimental timescale.
We therefore solve:
\begin{equation}
\label{eqn:ohmic_heating}
\frac{3}{2}\frac{\partial{}n_{e}T_e}{\partial{}t}=\eta_{Sp} j^2
\end{equation}
using the Spitzer-Braginskii resistivity \(\eta_{Sp}\propto{}T_e^{-3/2}\) and the current density shown in \hyperref[fig:faraday]{Fig. 3c}, and find that the time to heat electrons from 15 eV to 100 eV is \(\tau_{res}=350\) ns.
This time scale is also too long --- Spitzer-Braginskii resistivity cannot significantly heat the electrons during this experiment.

In other experiments, anomalous resistivity \cite{Ji2004} and viscosity \cite{Hsu2001a} --- as might arise from particle scattering from waves driven by, for example, the lower-hybrid drift or the ion-acoustic instabilities --- have been invoked to explain high ion or electron temperatures.
In our experiments we observe \(T_i\approx\bar{Z}T_e\) and \(C_{i,A}\approx{}u_{ed}\) (\(u_{ed}\) is the electron drift velocity, \(j=en_{e}u_{ed}\)) which are common criteria for the development of such instabilities.
The presence of kinetic instabilities will be investigated using Thomson scattering in future experiments.

\begin{figure}[!h]
\label{fig:fastframe}
\ifpdf
\includegraphics[scale=1]{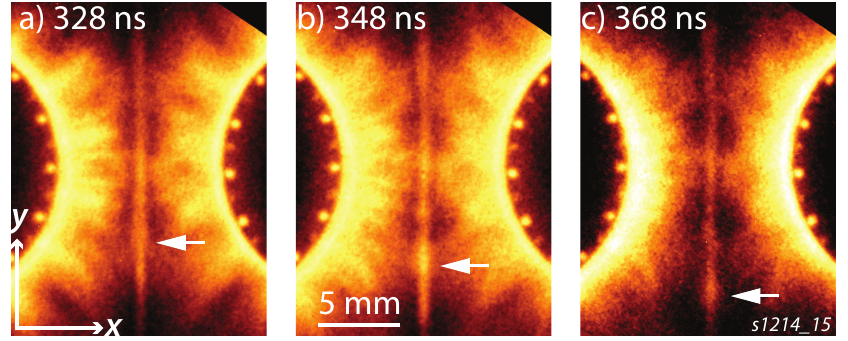} 
\else
\includegraphics[scale=1]{reconnection_paper-05.eps} 
\fi
\centering
\caption{Plasmoid formation and dynamics in three optical self emission images from the same experiment, 5 ns exposure, 20 ns between frames. The location of one plasmoid in each frame is indicated with a white arrow.}
\end{figure}

Another possible explanation for the anomalously high ion and electron temperatures is the plasmoid instability.
We observe plasmoids in electron density maps (\hyperref[fig:density]{Fig. 2a \& b}), and multiple plasmoids in fast-frame optical self emission imaging (\hyperref[fig:fastframe]{Fig. 5}, and also in the supplementary video \url{https://goo.gl/OjqA4M}).
A tentative explanation is that our experiment, with \(S=120\) and \(L/d_i=18\), sits in the semi-collisional regime of the plasmoid instability (eqn. 5, \cite{Loureiro2015}).
The plasmoid instability breaks the current sheet into numerous smaller sheets; in MHD, this is known to enable the rapid and efficient conversion of magnetic energy to thermal and kinetic energy \cite{Loureiro2012}.
It is unknown whether this enhanced heating should be observed in the semi-collisional regime.

Plasmoids have recently been observed in experiments on TREX \cite{Olson2016} and MRX \cite{Jara-almonte2016}, but in a parameter regime in which no plasmoids are predicted to form.
In contrast, in the semi-collisional regime the theoretical linear growth time of the plasmoid instability is \((L/d_i)^{6/13}S^{-7/13}L/V_A\sim30\) ns and the number of plasmoids predicted is \((d_i/L)^{1/13}S^{11/26}/2\pi\sim3\) \cite{Baalrud2011}.
This growth rate is consistent with the presence of plasmoids in this experiment, as the instability could grow on the experimental time-scale.
The number of plasmoids expected in the linear regime (unresolved in this experiment) is consistent with the number we resolve in the non-linear regime.
Our results thus open up the study of plasmoids in a new and distinct region of reconnection parameter space.

In summary, we have presented the first experimental evidence for magnetic reconnection in a pulsed-power driven experiment in which \(\beta_{dyn}\sim\beta_{th}\sim 1\) and \(M_A\sim 0.7\). 
Colliding flows produce a well-defined, large aspect ratio reconnection layer, which persists for more than ten hydrodynamic crossing times.
In this layer we observe the annihilation of magnetic flux and the acceleration and heating of the plasma.
Compressibility and pressure balance effects explain the fast inflows and outflows, and the measured power flowing into the layer is well matched by the measured power flowing out.
The ion and electron temperatures are anomalously high, with the ion temperature significantly larger than the electron temperature.
These high temperatures may be due to the plasmoid instability that we observe or, alternatively, to anomalous resistivity and viscosity triggered by kinetic instabilities.

This work was supported in part by the Engineering and Physical Sciences Research Council (EPSRC) Grant No. EP/N013379/1, by the U.S. Department of Energy (DOE) Awards No. DE-F03-02NA00057 and No. DE-SC-0001063, and by the LABEX Plas@Par with French state funds managed by the ANR within the Investissements d'Avenir programme under reference ANR-11-IDEX-0004-02.

\bibliography{library}

\end{document}